# Case Study Based Software Engineering Project Development: State of Art

Sukhpal Singh[#1], Harinder Singh[*2]

[#1,*2] Computer Science and Engineering Department, Thapar University, Patiala, India

[1]ssgill@hotmail.co.in

[2]singh.harinder@outlook.com

*Abstract*— **This research paper designates the importance and usage of the "case study" approach effectively to educating and training software designers and software engineers both in academic and industry. Subsequently an account of the use of case studies based on software engineering in the education of professionals, there is a conversation of issues in training software designers and how a case teaching method can be used to state these issues. The paper describes a software project titled Online Tower Plotting System (OTPS) to develop a complete and comprehensive case study, along with supporting educational material. The case study is aimed to demonstrate a variety of software areas, modules and courses: from bachelor through masters, doctorates and even for ongoing professional development.**

*Keywords*— **Software Engineering, Case Study, Software Project Management, Software Designer, Software Engineer**

## I. INTRODUCTION

Case study technique was first presented into social science by Frederic Le Play in 1829 as a handmaiden to indicators in his studies of family financial plan [1]. Case studies are studies of persons, actions, choices, stages, projects, strategies and organizations are considered holistically by one or more approaches [2]. The situation that is the subject of the review will be an instance of a class of occurrences that offers a systematic structure; an object, within which the revision is conducted and which the incident irradiates and elucidates [3]. It is a research approach, an experimental analysis that scrutinizes an occurrence within its real life perspective [4]. Case study research can mean single and various case studies can contain measureable substantiation, trusts on several sources of substantiation, and paybacks from the previous development of hypothetical propositions [5]. Case studies must not be chaotic with qualitative research and they can be based on any mix of reckonable and qualitative proof [6].

Software engineering (SE) is the application of a methodical, well-organized, countable method to the strategy, development, procedure, and maintenance of software, and the revision of these methods; that is, the application of engineering to software [7]. The usually acknowledged thoughts of SE as an engineering discipline have been identified in the guide to the Software Engineering Body of Knowledge (SWEBOK) [8]. The SWEBOK has become a worldwide recognized standard ISO/IEC TR 197 59:2005. Case studies and illustrations support you recognize real systems and demonstrate some of the applied complications of SE [9]. A Software Development Life Cycle (SDLC) is fundamentally a sequence of phases that offer a model for the development and lifecycle management of software [10]. The procedure within the SDLC process can fluctuate through industries and administrations, but standards such as ISO/IEC 12207 signify processes that inaugurate a lifecycle for software, and provide an approach for the development, acquirement, and configuration of software systems [11]. SDLC in SE, information systems and SE is a process of generating or modifying information systems, and the models and approaches that people use to develop these systems [12]. In SE, the SDLC model reinforces various types of software development practices [13]. These policies form the framework for forecasting and monitoring the establishment of an information system i.e. the software development process [14].

Software Project Management (SPM) is the talent and science of development and leading software projects [15]. It is a sub-discipline of project management in which software projects are intended, realized, supervised and organized [16]. The rest of the paper is structured as follows. Case Study based Project Development has been presented in Section 2. In Section 3, a description of the Problem Identification has been presented. Project Charter has been presented in Section 4. Section 5 describes the Scope Management. Section 6 presents the Time Management. Cost Management has been presented in Section 7. In Section 8, Quality Management has been presented. Communication Management has been presented in Section 9. Section 10 describes the Risk Management. Conclusions and the future works have been presented in Section 11.

## II. CASE STUDY BASED PROJECT DEVELOPMENT

Case studies were first used in the Harvard Law School in 1871 [4]. Meanwhile then, case studies have been a subject of much study and research about their usefulness in teaching and learning [5]. They have become a supported and ubiquitous way of teaching about professional training in such fields as commercial, law, and medicine. In its most simple form, it merely denotes to a genuine illustration used to demonstrate a thought. More formerly, and for the dedications of the Case Study Task, a case study encompasses the application of understanding and abilities, by an individual orgroup, to the identification and answer of a problem related with a realistic situation. Such a case study would comprise an account of a practical activity, occurrence, or circumstances. It might be supported with background material (setting, qualities, series of actions, and difficulties and clashes), objects, and data, which is related to the situation portrayed. Case studies are intended to inspire contribution, discussion and understanding. Though they can be used in an informative, teacher-cantered pedagogics, they are most often used in a dynamic learning, student-cantered system where the instructor acts as an originator or instructor. Case studies are of distinctive value in problem-based learning, which focuses on the development of problem-solving skills, self-directed learning abilities, and team skills [6]. In our project, we realize a case study as a means to pretend "evolutionary approach" and to expose students to the disorder and unrestrained nature of real-life issues.

A. *Purpose of Case Study*

This research paper delivers an explanation of a current software engineering project in the Computer Science and Engineering Department at Thapar University, Patiala, India. Unceremoniously it is called the "Case Study Mission". The project team includes of an Associate Professor of software engineering and two graduate students in the Master of Software Engineering program. The preliminary stage of the project, described in this paper, lasted for six months, in 2011. The Case Study project has the same determination, with the following more specific goals:

- Create software development items, which offer a "practical" foundation for teaching SE.
- Design and consolidate the real-world software development objects into a set of case modules (mini case studies), which can be used during a computing program.
- Deliver SE resource materials, which can be adapted to several teaching and learning styles and methods.
- Involve software engineering gurus in the valuation, use, and development of the case study materials.

The shorter term project objective was to create anagenda for support of a full case study, which wouldinclude the following activities:

- Establish the project, process and planning procedures.
- Conduct research into case study lessons.
- Determine a problem area and select an application in that area, which would be the subject of the case study.
- Create a basis situation for the case study application (e.g., explanation of real-life circumstances that founds the need for the case application, presents people who will be part of the case study, and presents conditions that will restrain and guide the application development).

B. *Problems in SE Project Development*

Teaching SE in a bachelor and master's program has two main obstacles:

- SE is a fresh and evolving discipline. It is not so far advanced and some even query whether it is engineering. This partially describes the shortage of material to support teaching SE. The ACM/IEEECS Recommendations on SE Teaching [17] should support advance and sustenance not only the teaching of SE, but the development of critical support materials (e.g., workbooks and web resources).
- SE is a proficient field and scholars need more than courses in basics and theory; they want to do study about and experience professional practice. One of the prospectus strategies in [17] states that "The prospectus should have an important real-life foundation". Another guideline states "SE theories, philosophies, and issues must be accomplished as recurring refrains during the prospectus to help scholars develop a SE mentality". The question is how to superlative provide this "real-life" experience and explain "recurring themes" that develop a "SE mentality".

C. *Case Study Solution*

The Case Study Project (OTPS) focuses on developing case modules, which are associated by being part of and derived from a single case, the development of a single software product. Each case module relates to a movement involved in the development of the product. In addition, each case component is outlined as part of a product development description, using a scenario format, which involves characters and occurrences that might be part of a real SPM (e.g., creation of a software project team, communication with upper management, customer and customer meetings, Creation of Problem Identification, Project Charter, Scope Management, Time Management, Cost Management, Quality Management, Communication Management, Risk Management., etc.). Though the case study materials focus on one domain area and one development method (process and practice), it can also assist as an illustration of how to develop related case studies, using other domain areas and other development methods. A case module could be considered a "mini-case study".

The Case Study Project is intended to cover the complete SDLC (project management, requirement analysis and specification, design, implementation, testing and maintenance). For the preliminary phase of the case study project it was decided to focus on constructing a basis for complete development: research into case study teaching; recognize a case study problem; describe a launch of the software development team; develop a software development plan to be used as part of the case study; and develop numerous related case components.

D. *Case Study Elements*

The long-term plan is to write approach, develop OTPS artifacts, and create processes, data and reports that pretend all of the development team's work on the OTPS project. Thus far, three phases (Project Inception, Project Launch, and Planning) have been accomplished. In these phases the following OTPS case study documents were developed:

Approaches
- Inception Approach
- Team Real Drafts
- Launch Approach
- Planning Approach

OTPS Artifacts
- Project Charter
- Scope Statement
- Software Requirement Specification
- Data Flow Diagram
- Data Dictionary
- UML Diagrams
- Work Breakdown Structure
- Time Management
- Cost Management
- Quality Policy
- Quality Management Plan
- Quality Review
- Test Plan
- Test Case
- Test Result
- Control Flow Graph
- Linearly Independent Path
- Audit Report
- Communication Management Plan
- Stakeholder communication Analysis
- Risk Management Plan

### III. PROBLEM IDENTIFICATION

The project team considered and studied a number of possibilities for the case study problem: a computer game, air traffic control software, an academic planner, a math education application, an immune system simulation, a weather reporting system, and the software system for "Plotting towers". After consultation with colleagues teaching software engineering at other schools and interactions with undergraduate and graduate software engineering scholars, the team decided that the best candidate for the case study was a map generator. For the purposes of the Case Study project, a Smart Map is defined as an OTPS that allows plotting towers on the defined map. After some additional research on Graphical Information System (GIS) technology, including a discussion with a Reliance Telecom, the team settled on development of a software system that would support a smart tower plotting system called OTPS. This Project will be design for the Telecom Company. It helps in plotting the tower images on Satellite Map. Site engineer can select particular shape and colour of the mark that will reflect the position of a particular tower on the satellite map. It holds the information about all the towers of operators in various site locations which control different MSCs in company and getting tower information by clicking on the image. In this we can create, delete and update the tower information. This software system helps in calculating the distance between two towers in a particular MSC. The search option will be there for searching a particular tower by name, site identity number, latitude and longitude. For adding information about a particular site, the Site engineer can upload the file in a format supported by the software, which is the indirect mean of adding information regarding new sites/towers into the database. For security reasons this project provides usernames and passwords to control the information. Also a site engineer can create new login ids for his/her subordinates (like trainee etc.) who can also manage the sites in concurrently. The user can move to any location by dragging the mouse over the Satellite map window and user can also zoom in and zoom out by using zoom slider.

### IV. PROJECT CHARTER

Project Charter is a statement of the scope, purposes and contestants in a project. It deliversaninitialexplanation of characters and accountabilities, frameworks the project aims, identifies the keyparticipants, and describes the power of the project manager. It works as a reference of power for the future of the project.

*A. Project Charter Attributes*
- Project Title: Online Tower Plotting System.
- Project Start Date: 2 August, 2011
- Project Finish Date: 10 November, 2011
- Budget Information: An initial estimate provides a total 18 hours per week including lab work.
- Project Manager: Dr. Inderveer Chana, Associate Professor (CSED, Thapar University, Patiala).

*B. Project Objectives*

Develop an automated OTPS for Telecom Company helps site engineer to manage the information about towers, sites and users effectively. OTPS manages the various types of networks like GSM and CDMA. The different mobile switching controls are allocating to different site engineers to maintain the whole database. OTPS is replacing the existing in which information is managing manually.

*C. Software Development Approach*
- Develop a survey in various telecom companies to determine the important features of OTPS.
- Review the documents of other similar systems relating to the OTPS.

- Develop online tower plotting system using Rapid Application Development.
- Collecting the whole information about the towers and sites from Telecom Company.

*D. Software Development Team*

The Roles and Responsibilities of OTPS are shown in Table 1.

TABLE I
ROLES AND RESPONSIBILITIES

| Name | Role | Position | Contact Information |
|---|---|---|---|
| Dr. Inderveer Chana | Supervisor and Instructor | Project Manager | inderveer@thapar.edu |
| Harinder Singh | Team member | Developer | singh.harinder@outlook.com |
| Sukhpal Singh | Team member | Developer | ssgill@hotmail.co.in |

V. SCOPE MANAGEMENT

Scope management is significant for operational project management. Projects are anticipated to fulfilstringentgoals with resource restraints, and an unvented and unapproved change in the scope can affect the success of the project. Scope sometimes causes cost overrun.

*A. Scope Statement*

Scope statements may take many forms depending on the type of project being implemented and the nature of the association. The scope statement specifics the project deliverables and designates the chiefintentions. The intentions should comprisequantifiable success standards for the project.

- Project Justification: An automated global positional system: OTPS for mobile operator company helps site engineer to manage the information about towers, sites and users effectively. It facilitates the job of site engineer by automatically placing towers graphically. OTPS manages the various types of networks like GSM and CDMA. It reduces the workload of site engineers, no need to maintain the data registers manually. The different mobile switching controls are allocating to different site engineers to maintain the whole database. It needs just one time efforts while developing the complete product, after that site engineer can maintain and perform number of functions on it.
- Product Characteristics and Requirements: It includes main features of OTPS: Graphical user interface, Links, Security, Plot tower images, Signup, Search feature, Operations (view, update, delete and add new tower) Calculate distance, Uploading data input, Availability, Zooming.
- Detail of project deliverables: Project management related deliverables: Project charter, scope statement, SRS, WBS, Cost baseline, final project report and any other documents required to manage the project.

*B. Software Requirement Specification*
1) Introduction
    - Purpose: This document is intended for customers from various Telecom Companies for which this software product is being developed, Site engineers of the company, developers, marketing managers, software testers, software architect. This document specifies various functional as well as non-functional requirements in detail and it describe which functional/non-functional requirement are satisfied by which feature of the software.
    - Document Conventions: OTPS - Online Tower Plotting System etc.
    - Intended Audience and Reading Suggestions
    - OTPS Scope
    - References
2) Overall Description
    - OTPS Perspective
    - OTPS Features
    - User Classes and Characteristics
    - Operating Environment
    - Design and Implementation Constraints
    - User Documentation
    - Assumptions and Dependencies
3) System Features
    - Search
    - Tower Definition
    - Distance Calculation
    - File Upload
4) External Interface Requirements
    - User Interfaces
    - Hardware Interfaces
    - Software Interfaces
    - Communications Interfaces
5) Other Nonfunctional Requirements
    - Performance Requirements
    - Safety Requirements
    - Security Requirements
    - Software Quality Attributes
6) Other Requirements

*C. Data Flow Diagram*

A data flow diagram (DFD) is a graphical demonstration of the "flow" of data through an information system, modelling its process features. Frequently they are a primary step used to create an outline of the OTPS which can later be expanded. The 0 level or context flow diagram of OTPS is shown in Figure 1. This context-level DFD is next "detonated", to yield a Level 1 DFD that shows some of the detail of the OTPS being modelled.

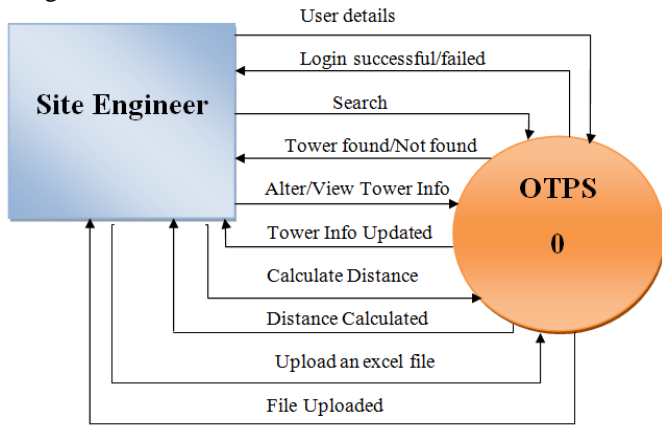

Fig.10 Level or Context Level DFD

The Level 1 DFD shows how the Online OTPS is distributed into sub-systems (processes), each of which deals with one or more of the data flows to or from an external agent, and which together offer all of the functionality of the OTPS as a whole. It also recognizes internal data stores that must be existent in order for the OTPS to ensure its work, and displays the flow of data among the many parts of the OTPS. The internal data stores in OTPS are user database and site database. User database manage all the information of existing users while site database mange the tower and site information. The 1 Level DFD is shown in Figure 2.

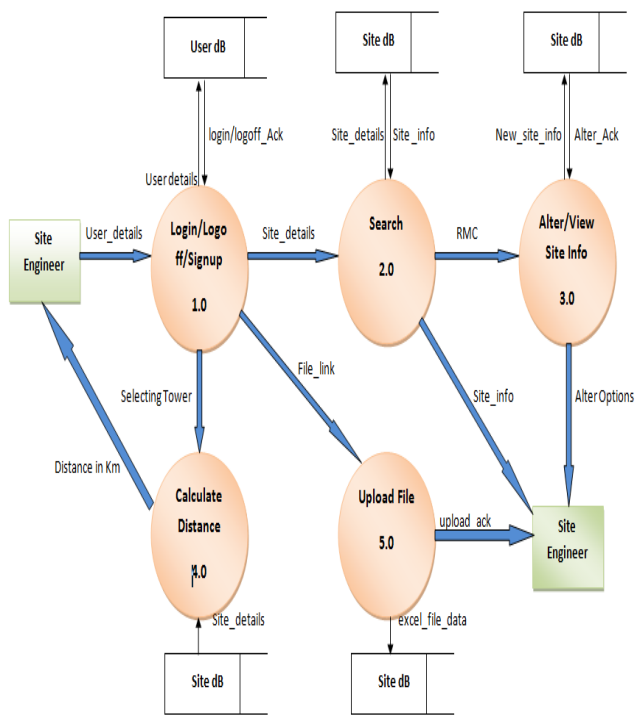

Fig. 2 1st Level DFD

The 2 Level DFD is shown in Figure 3. In this Figure user Login, Signup, Logoff functions are explained and flow of data from one process is explained in 2 Level DFD.

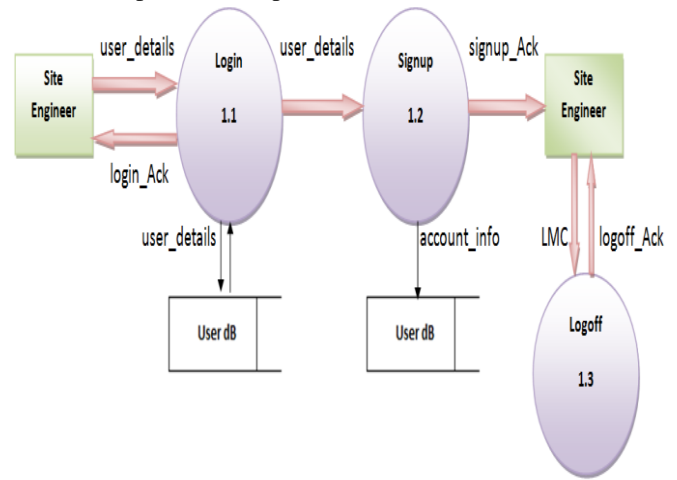

Fig. 3 2nd Level DFD

*D. Data Dictionary*

A data dictionary, or metadata warehouse, is a compacteddepository of information about data such as denotation, relationships to other data, origin, usage, and format. The data dictionary (Databases) of OTPS is described in Table II.

TABLE II
DATA DICTIONARY

| Data store name | Description | Inbound data flow | Outbound data flow |
|---|---|---|---|
| User Database | All the information of site engineer are stored in the user database, when the site engineer login then in compare the user detail with data store in user database. | Login/ Signup | Login/log out/signup |
| Site database | All the information of all the sites is stored in the site database in an appropriate form. | Search, Alter/view site info, Upload file, Calc_dist | Search, Alter/view site info, Calc_dis |

*E. UML Diagrams*

UML (Unified Modelling Language) is a graphical language for envisaging, specifying, creating and documenting the artifacts of OTPS. The UML gives a standard way to write an OTPS's blueprint, covering conceptual things, such as business processes and OTPS functions as well as such as classes written in a particular Programming Language, databases schema and reusable software components. In the design of OTPS, we have developed eight UML diagrams: Use case diagram, Class diagram, Activity diagram, Sequence diagram, Collaboration diagram, State chart diagram, Component diagram and Deployment diagram.

1) *Use Case Diagram*: The use case diagram is essential modelling the behaviour of the OTPS, subsystem or classes. The interaction among use cases and actors is described in this diagram. The use case diagram of OTPS is shown in Figure 4.

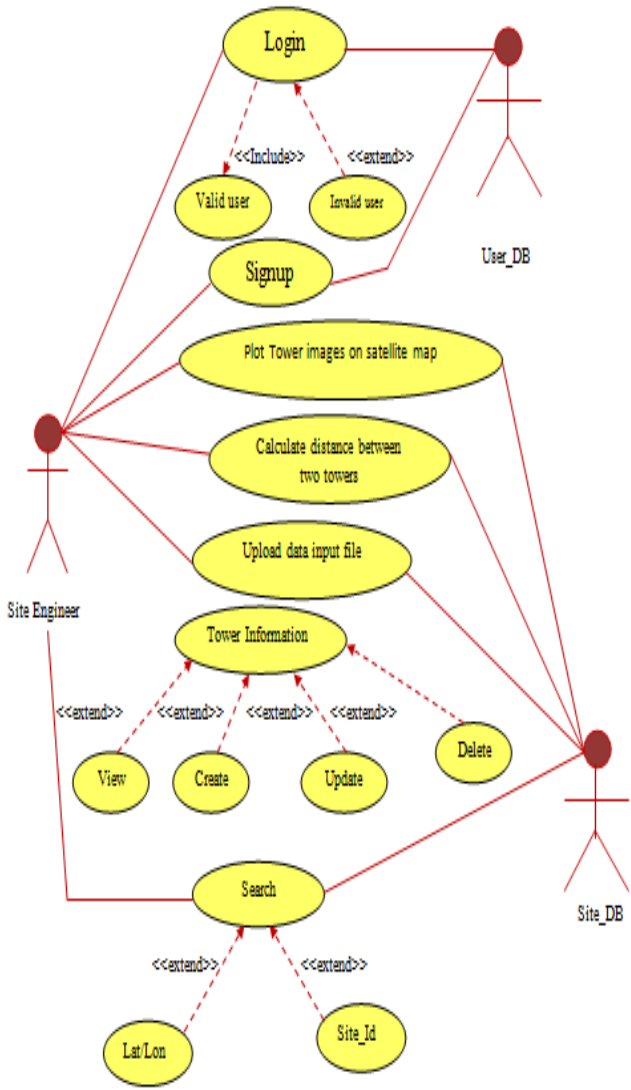

Fig. 4 Use Case Diagram of OTPS

2) *Class Diagram:* A class diagram shows a collection of classes, interfaces, associations and their interactions. The class diagram of OTPS (Login) is shown in Figure 5.

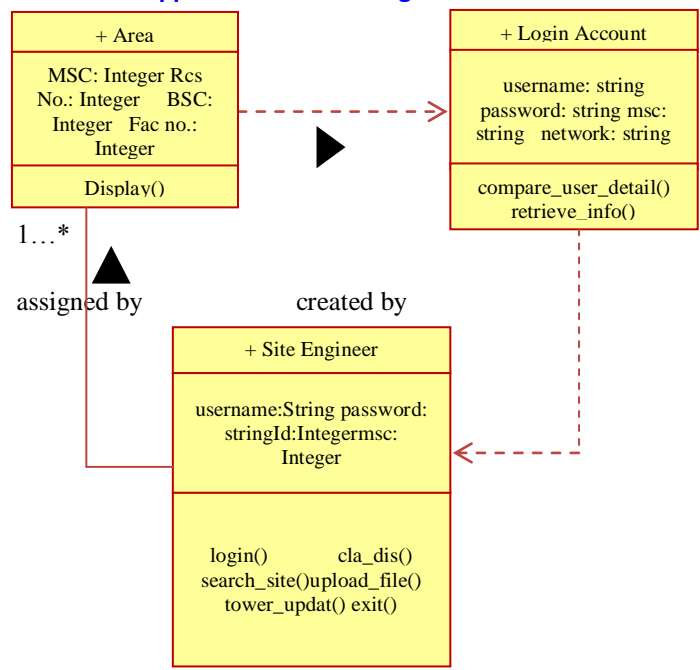

Fig. 5 Class Diagram of OTPS

3) *Activity Diagram:* Activity diagrams are graphical representations of workflows of stepwise activities and actions with sustenance for selection, repetition and concurrency. Activity diagrams show the overall flow of control. The activity diagram of OTPS is shown in Figure 6.

4) *Sequence Diagram:* A sequence diagram is a kind of interaction diagram that shows how processes operate with one another and in whateverdirection. A sequence diagram displaysitemcollaborationsorganized in time order. It represents the objects and classes used in the scenario and the sequence of messages exchanged between the objects needed to carry out the functionality of the scenario. The sequence diagram of OTPS (Login) is shown in Figure 7.

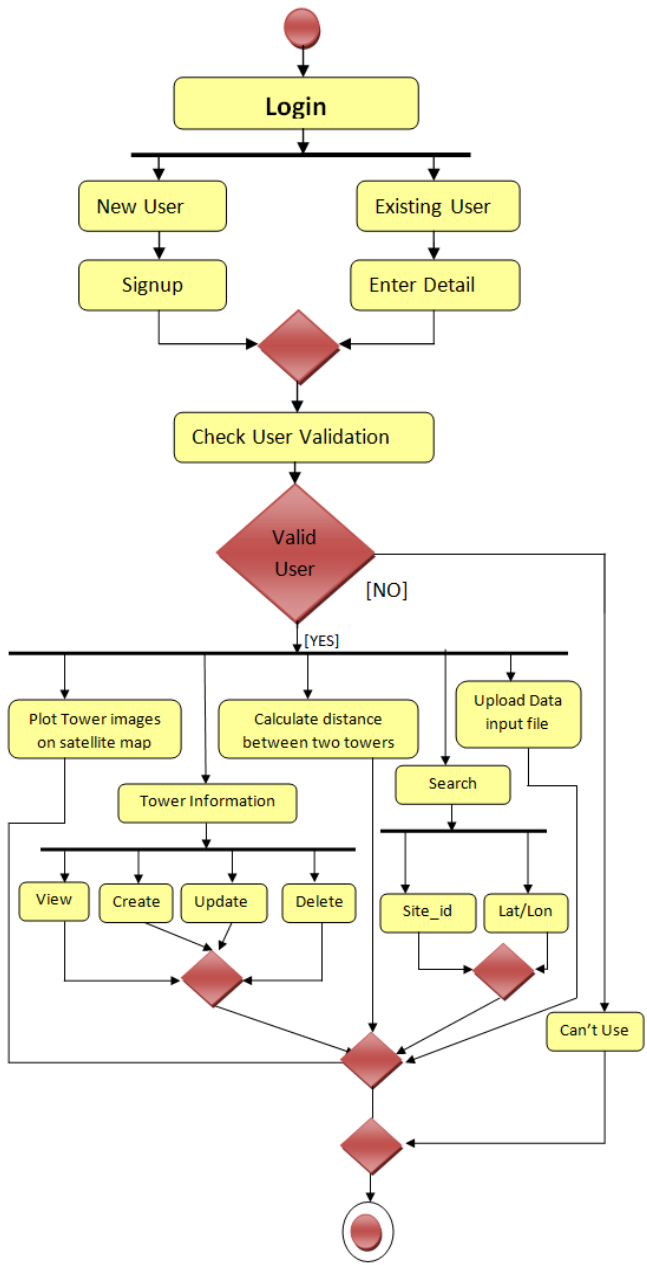

Fig. 6 Activity Diagram of OTPS

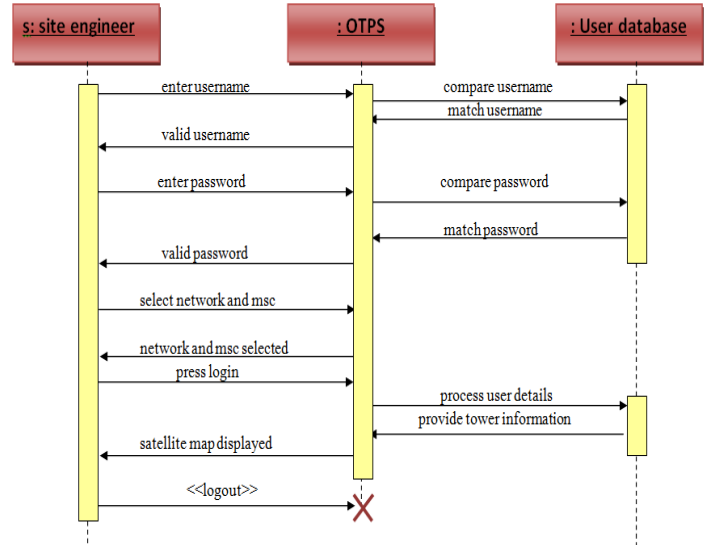

Fig. 7 Sequence Diagram of OTPS

*5) Collaboration Diagram:* Collaboration diagrams represent a combination of information engaged from Class, Sequence, and Use Case Diagrams defining both the static structure and dynamic behaviour of anOTPS.The collaboration diagram of OTPS (Login) is shown in Figure 8.

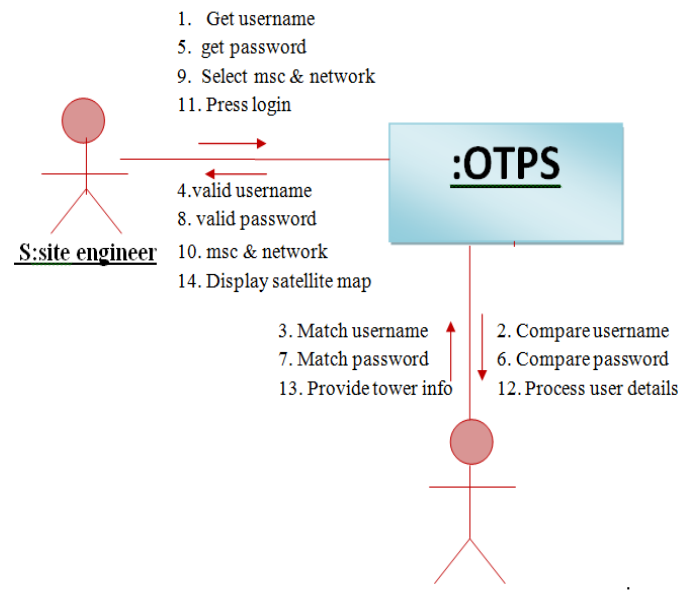

Fig. 8 Collaboration Diagram of OTPS

*6) State Chart Diagram:* State cart diagrams need that the OTPSdesignated is composed of a predictable number of states; occasionally, this is certainly the case, whereas at other times this is a realistic abstraction. The state chart diagram of OTPS is shown in Figure 9. There are three things that are required by every state for its transition: 1) Function 2) Condition and 3) Action. States in state chart diagrams represent a set of those value groupings, in which an object performs the similar in comeback to actions. This diagram models the dynamic flow of control from state to state within an OTPS.

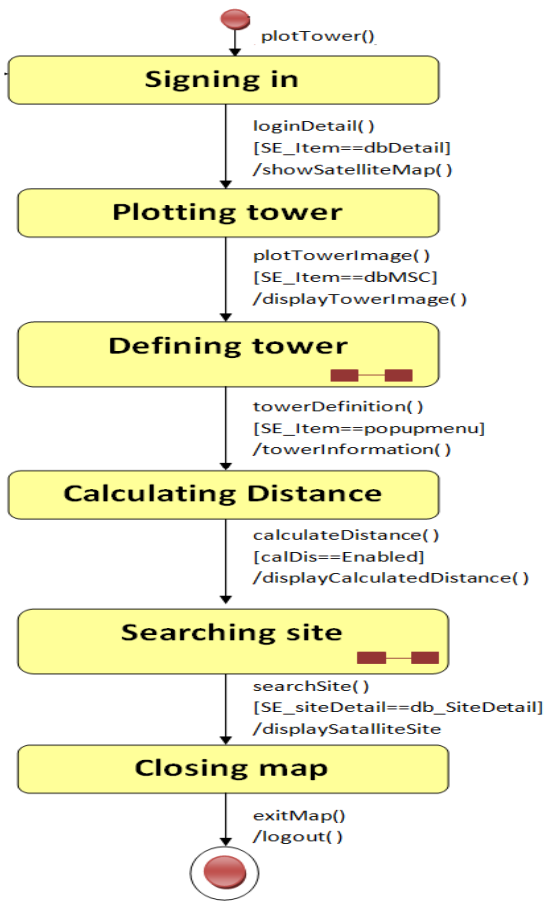

Fig. 9 State Chart Diagram of OTPS

The detailed description of state "Defining Tower" is described in Figure 10 as a separate state.

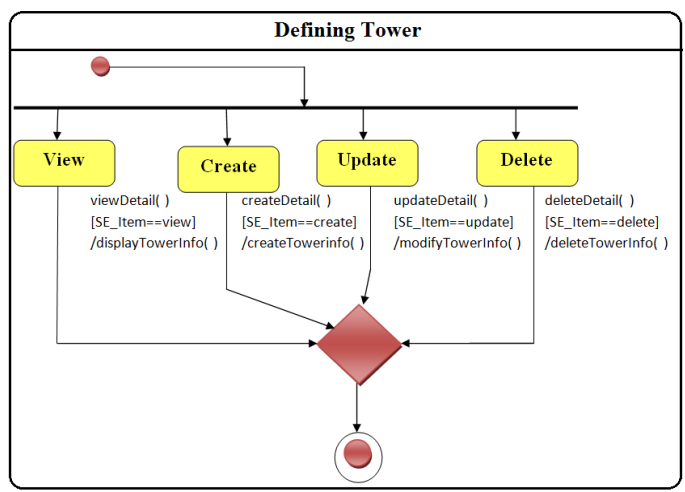

Fig. 10 Defining Tower State

*7) Component Diagram:* Component diagram depicts how components are wired together to form larger components and or software systems. Components are held together by using an assembly connector to connect the required interface of one component with the provided interface of another component. The component diagram of OTPS is shown in Figure 11.

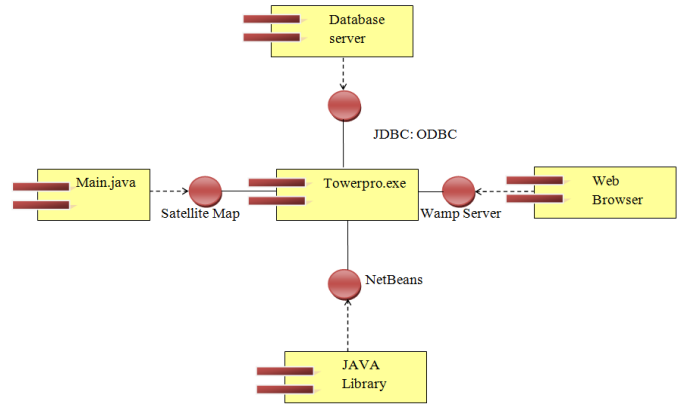

Fig. 11 Component Diagram of OTPS

Modelling Executable and Libraries associated with the above Components are described in Figure 12.

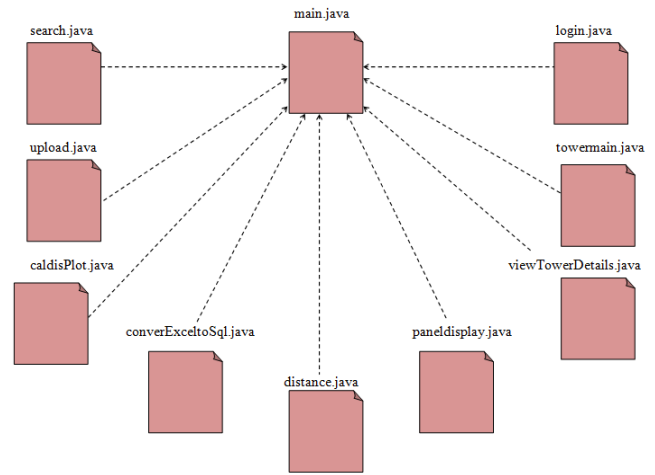

Fig. 12 Modelling of Executables and Libraries

*8) Deployment Diagram:* Deployment diagram models the physical deployment of artifacts on nodes. The component diagram of OTPS is shown in Figure 13.

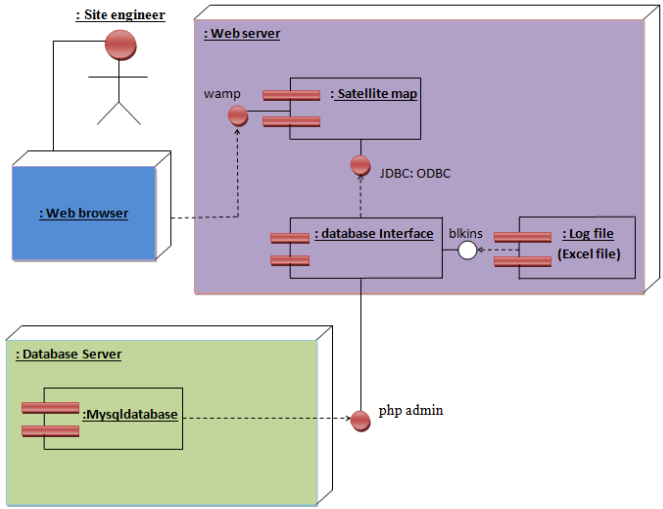

Fig. 13 Deployment Diagram of OTPS

A deployment diagram would demonstratewhatever hardware components (nodes) are, what software components (artifacts) run on each node (Web server, database server and web browser), and how the different pieces are connected.

*F. Work Breakdown Structure*

A work breakdown structure (WBS), in project management and systems engineering, is a deliverable basedbreakdown of a project into minor components. It describes and clusters a project's isolated work elements in a way that helps organize and define the total work scope of the project. The phase based WBS is shown in Figure 14, it explains all the phases of SDLC.

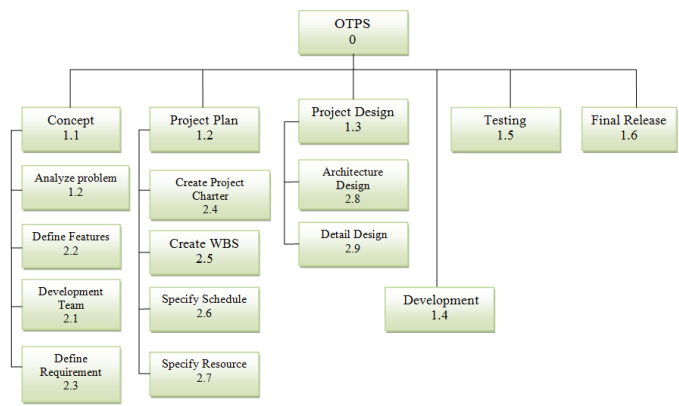

Fig. 14 Phase Based WBS

Based on the features delivered in the OTPS, the product based WBS is shown in Figure 15.

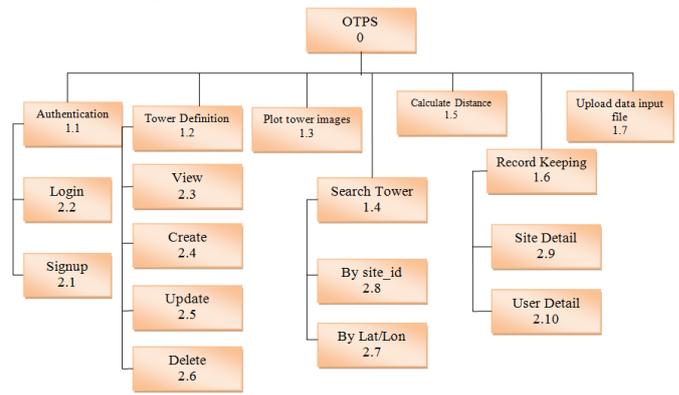

Fig. 15 Product Based WBS

VI. TIME MANAGEMENT

.The objective of time management is how to manage time and why managing time is important. The time taken by every activity is measured by network diagram.

*A. Network Diagram*

Network diagram are the preferred technique for showing activity sequencing. A network diagram is a semantic display of logical relationships among sequencing of project activities. The network diagram of OTPS is shown in Figure 16.

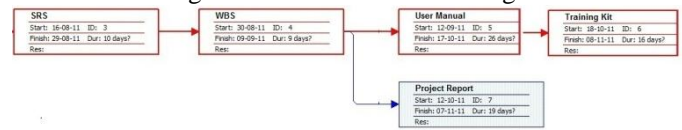

Fig. 16 Network Diagram of OTPS

VII. COST MANAGEMENT

*A. Project Cost Baseline*

The project cost baseline is the foundation for the earned value recordingscheme. It is the financial plan for the projectedbudget of the project spread over the time periods of the project.The project cost standard is the portion of the project concerned with the quantity of money that the project is forecast to charge and when that cash will be used. The three standards are thoroughlylinked, and variations to one of them will result in changes to the others. If a modification is made in the project scope baseline, either by adding or eliminatingcertain of the effort that is mandatory, the timetable baseline and the cost baseline will probably have to be altered as well. The project cost baseline of OTPS is shown in Table III.

TABLE III
PROJECT COST BASELINE

| Task | Resources | Effort Estimates (Days) | Rate | Total (Rs.) |
|---|---|---|---|---|
| **Requirement Gathering** | | | | 1300/- |
| Travelling | Transportation | 2 | 200/ | |

| Meeting | Shared Resources: Library Staff Restaurant | 3 | 300/- | - |
|---|---|---|---|---|
| **Data Collection** | Supporting Organization | 1 | 100/- | 100/- |
| **Learning and Training** | Professional SE Community | 10 | 25/- | 250/- |
| **Documentation** | Computer Resources | 5 | 60/- | 300/- |
| **Lab Work** | | | | 1210/- |
| Using Tools | Software tools | 10 | 45/- | |
| Coding | Programming Environment | 13 | 20/- | |
| Using Support Software | Internet | 3 | 50/- | |
| Networking and communication | Hardware | 7 | 50/- | |
| **Total** | | | | **3160/-** |

*B. Resource Allocation Matrix*

The resource allocation matrix structure has become the primary organizational means for maintaining an efficient flow of resources in multi-project environments. The resource allocation matrix used for OTPS is shown in Table IV.

TABLE IV
RESOURCE ALLOCATION MATRIX

| Activities / Resources | C | PP | PD | D | T | R |
|---|---|---|---|---|---|---|
| **Internet** | | | | | √ | √ |
| **Computer** | | | √ | √ | √ | √ |
| **Software** | | | √ | √ | √ | √ |
| **Hardware** | | | √ | √ | √ | √ |
| **Electricity** | | | √ | √ | √ | √ |
| **Transportation** | √ | √ | | | | |
| **Human** | √ | √ | √ | √ | √ | √ |
| **Supporting Organization** | √ | √ | | | | |
| **Data Source** | | | | | √ | √ |
| **Financial** | √ | √ | √ | √ | √ | √ |

The abbreviations used in resource allocation matrix are C: Concept, PP: Project Pan, PD: Project Design, D: Development, T: Testing, R: Rollout. Vertical rows denote resources and horizontal columns denote resources.

VIII. QUALITY MANAGEMENT

The aim of Quality Management (QM) is to manage the quality of software and of its development process. A quality product is one which meets its requirements and satisfies the user. A quality culture is an organizational environment where quality is viewed as everyone's responsibility.

*A. Quality Policy*

We are committed to develop a high quality, technical and engineered software product, which is efficiently adoptable by different multinational organizations whosoever, works in the telecom operator domain. This software is used by the site engineer of telecom operator organization and provides technical support to the engineer and assists him/her in facilitating their work, which was quite tedious to be managed manually. Thus it provide support to their job such that they can efficiently, precisely and confidently configure and manage their respective sites and have a confidence their useful and important information is secure and nobody as intruder can exploit and corrupt the confidential information. We have followed ISO-9126 standard practices for the development of this software system for quality aspect. We follow Rapid Application Development (RAD) software process model because of stringent time deadlines and we have fixed the project scope early before the start of the project.

*B. Quality Management Plan*

A Quality Management Plan documents how an association will design, implement, and evaluate the usefulness of its quality assurance and quality control processes. Precisely, it pronounces how an organization constructs its quality system, the quality policies and procedures, regions of application, and characters, accountabilities, and specialists.

- Product Introduction: This OTPS will be design for the Mobile Operator Company. It helps in automatic plotting of tower on Satellite Map. It holds the information about all the towers using in this company and getting tower information by clicking on the image. In this we can create, delete and update the tower information. This project helps to calculate the distance between two towers etc.
- Product Plan: The OTPS is of GPS domain software system, it is helpful in replacing the system in which we collecting the whole information regarding towers of telecom company in a particular site location with a fully

ISSN 2319 – 1953International Journal of Scientific Research in Computer Science Applications and Management Studiesautomated system in which we can do all these activities on online map, we convert manual tower searching system in registers to a fully automated system. This software system is intended for different telecom or mobile operator companies.

- Process Descriptions: The software team selects a process for the work to be performed. The project manager must decide which process model is most appropriate based on the customers who have requested the product and the people who will organize the work, the features of the product itself, and the project situation in which the software group works.
- Quality Goals: This system is adaptability any file containing information of site details can be inserted to the database. It's availability is high, it depends only on the maximum number of internet users allowed, it's correctness is high, system provides the exact identification of the found tower on the graphical map by showing it in a particular colour and shape which is set by user before doing a search. It also has some more quality attributes like reliability, reusability, testability, flexibility, security and usability.
- Risks and Risk Management: It is necessary to anticipate and identify different risks that a project may be susceptible to, so contingency plans can be prepared to contain the effects of each risk. Risk management aims at reducing the impact of all kinds of risks that might affect a project. It needs to anticipate the risks in the project as early as possible so that the impact of risks can be minimized by making effective risk management plans.

*C. Quality Review*

A process or meeting during which a software product is examined by project workforces, administrators, consumers, customers, user senates, or other interested parties for comment or approval. Different type of quality reviews are described in Table V.

TABLE V
QUALITY REVIEW

| Name | Purpose |
|---|---|
| SRS Review | A review of the Software Requirements Specification of OTPS is conducted by both the software developer and the Site engineer. Because the specification forms the foundation of the expansion phase, great care should be taken in directing the assessment. |
| Peer Review | Peer review is performed for all the documents and code of all the modules or functionalities of OTPS by the respective authors and developers respectively during different software development activities. It is done to provide a disciplined engineering practice for detecting and correcting the defects in software artifacts. |
| Formal Technical Review (FTR) | FTR is a software quality assurance activity performed by software engineers. The objectives of the formal technical review are to uncover faults in function, logic, or execution for any demonstration of the software, to authenticate that the software under analysis meets its requirements, to ensure that the OTPS has been represented according to predefined standards, to achieve OTPS that is developed in aidenticalfashion, to make projects more controllable. |
| Group Review | The SQA group appraisals the process explanation for acquiescence with organizational strategy, internal software criterions, superficiallycompulsory standards (ISO-9001), and other parts of the software project plan. The SQA group reviews selected work products: recognizes, documents, and tracks nonconformities; authenticates that improvements have been made; and occasionally reports the results of its work to the project manager. |
| Software Review | Analyses software engineering events to verify obedience with the well-defined software process. The SQA group recognizes, documents, and tracks abnormalities fromthe process and verifies that corrections have been made. OTPS reviews are a "filter" for the software engineering process. These reviewsare applied at various points during software development of OTPS and serve to uncover errorsand defects that can then be removed. |

*D. Test Plan*

The test plan for OTPS is described in Table VI. We have discussed only two types of testing here.

TABLE VI
TEST PLAN OF OTPS (LOGIN)

| Type | Name | Purpose | Performer |
|---|---|---|---|
| Unit Testing | | In order to test a single module, we will provide the complete environment to test each module independently like Login, Tower information, search, calculate distance and upload file. | Testing Team |
| Black Box Testing | | The test cases are designed from an examination of the input/output values only without considering code. | Site Engineer |
| | Equivalence Class Portioning | We perform testing of all the components based on respective classes of input. | Testing Team |
| | Boundary Value Analysis | We will perform the boundary value analysis of each input component of OTPS for different valid set of classes of input. | Testing Team |
| | | Designing white box test cases require a thorough knowledge of the internal structure of our project. | Tester |
| **White Box Testing** | Statement Coverage | This strategy aims to design test cases so that every statement in a coding part is executed at least once, to check the error if exist. | Test Engineer |
| | Branch Coverage | The test cases are designed to make each branch condition assume true and false in turn. | Tester |
| | Condition Coverage | Test cases are designed to make each component of composite conditional expression assume both true and false values. | Test Engineer |
| | Path Coverage | To design test cases such that all linearly independent paths in the OTPS are executed at least once. | Tester |
| | Cyclomatic Complexity Metric | This metric is used to define an upper bound on the number of independent paths in the OTPS. | Tester |
| | Data Flow Base Testing | This method selects the paths of the program according to the locations of the definitions and uses of different values in OTPS. | Tester |

The different roles assigned to people under a particular team for testing are:
- Testing Team: The team organized by Airtel and Vodafone Telecom Company.
- Test Engineer: Usability Engineer and Employee of Reliance Telecom Company.
- Site Engineer: Employee of Reliance Telecom Company.
- Tester: Harinder Singh and Sukhpal Singh

*E. Test Case*

A test case in software engineering is a set of conditions or variables under which a tester will determine whether anour application or software system is functioningproperly. The mechanism for defining whether a software program or system has passed or failed such a test is known as a test oracle. The test cases designed for testing of Login module of OTPS are described in Table VI.

TABLE VI
TEST PLAN OF OTPS (LOGIN)

| **Test Case: OTPS 1** | | **Screen Name: Login** | |
|---|---|---|---|
| Sr. No. | Input | Actual Output | Expected output | Result |
| 1. | Keep username blank | Username can't be blank | Invalid username | Pass |
| 2. | Keep password blank | Password can't be blank | Invalid password | Pass |
| 3. | Unselected MSB | MSB can't be unselected | Undefined MSB | Pass |
| 4. | Unselected network type | Network can't be unselected | Undefined Network | Pass |
| 5. | If either username or password is incorrect | Username and password can't be wrong | Invalid username and password | Pass |
| 6. | If both username and password is correct &MSB and network selected properly | It passes to Satellite Map | Valid username and password | Pass |

*F. Test Result*

The tester also created the results of the test execution, which are referred to as the test log.Test cases are executed by the tester and results of the tests are documented in the test log.he Test Analyst role is responsible for identifying and defining the mandatory tests, observingcomprehensive testing advancement and results in each test cycle and evaluating the overall quality experienced as a result of testing activities.The test cases designed for testing of Login module of OTPS are described in Table VII.

TABLE VII
TEST RESULT OF OTPS (LOGIN)

| **Input** | **Output** | **Consequences** |
|---|---|---|
| Username: admin<br>Password: admin<br>Msc: MSC 1 CHANDIGARH<br>Network: CDMA | Username, password, Msc, network art correct and it passes to satellite map | Satellite map shown |
| Username: spm<br>Password: spm123<br>Msc: MSC 5 JALANDHAR<br>Network: GSM | Username, password, Msc, network art correct and it passes to satellite map | Satellite map shown |
| Username: abc<br>Password: abc123<br>Msc: MSC 1 CHANDIGARH<br>Network: GSM | Username, password, Msc, network are incorrect and it asks site engineer to login again | Again login screen displayed |

*G. Control Flow Graph*

Path coverage based testing strategy requires us to design test cases such that linearly independent paths in the each module of OTPS are executed at least once. A linearly independent path can be well-defined in terms of CFG of OTPS. CFG describes the sequence in which the different instructions of OTPS executed. In order to draw CFG of OTPS, we need to first number all the statements of each module of OTPS. The different numbered statements serve as the nodes of CFG. An edge from one node to another node exists if the execution of statement representing the first node can result in the transfer of control to other node. The CFG for Login module is shown in Figure 17. We have drawn the CFG of some main modules of OTPS like:

- Login,
- Search site
- Upload file
- Calculate distance
- Tower definition

| Login |
|---|
| login_detail() { |
| 1. Enter user detail |
| 2. if (Enter_user_detail==db_user_detail) then |
| 3. Satellite map displayed |
| 4. else Enter user details again |
| 5. } |

The Cyclomatic complexity is calculates as given below:

| Cyclomatic Complexity : |
|---|
| **V(G)=E-N+2** |
| Edges (E) = 6 |
| Nodes (N) = 5 |
| V (G) = 6-5+2= 3 Or |
| **V (G) = Total number of bounded areas(regions) +1** |
| Regions = 2 |
| V (G) = 2+1=3 |

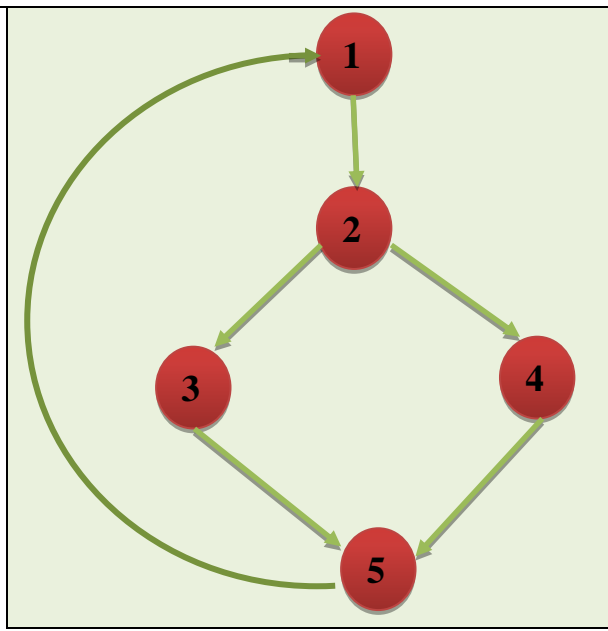

Fig. 17 Control Flow Graph of Login

*H. Linearly Independent Path*

A linearly independent path is a comprehensive path which, forgetting back tracking (such as loops), has a unique set of decisions in a program. It is any path through the program that introduces at least one new edge that is not included in any other linearly independent paths. If a path has one fresh node equated to all other linearly independent paths, at that time the path is also a linearly independent path. This is because whichever path having a new node spontaneouslyindicates that it has a new edge.The linear independent path of Login module of OTPS is described in Table VIII.

TABLE VIII
LINEARLY INDEPENDENT PATH

| Input | Path | Result |
|---|---|---|
| **Enter correct details:**<br>Username: admin<br>Password: admin<br>Msc: MSC 1 CHANDIGARH<br>Network: CDMA | **Path 1**:<br>1-2-3-5 | Satellite map shown |
| **Enter incorrect login details:**<br>Username: abc<br>Password: abc123<br>Msc: MSC 4 JALANDHAR<br>Network: GSM<br>**Enter again correct details:**<br>Username: admin<br>Password: admin<br>Msc: MSC 1 CHANDIGARH<br>Network: CDMA | **Path 2:**<br>1-2-4-5-<br>1-2-3-5 | Satellite map shown |
| **Enter incorrect login details continuously:**<br>Username: abc<br>Password: abc123<br>Msc: MSC 4 JALANDHAR<br>Network: GSM<br>………….. | **Path 3:**<br>1-2-4-5-<br>1-2-4-5-<br>1-2……. | Unsuccessful login |

*I. Audit Report*

The audit report is anofficialoutlook, or repudiation thereof, allotted by either an internal auditor or an independent external auditor as a result of an internal or external audit or evaluation performed on a legal entity or subdivision thereof (called an auditee). The report is subsequently provided to a "user" (such as an individual, a set of people, a corporation, a government, or even the common public, between others) as an assurance facility in order for the user to make decisions based on the results of the audit. OTPS is audited by two master students working on some other project. Both the auditors were assigned by our project supervisor. The audit report of OTPS is described in Table IX.

TABLE IX
AUDIT REPORT OF OTPS

| 1. Title of Project: Online Tower Plotting System |
|---|
| 2. Project member:-<br>Harinder Singh (Registration No.- 801131010)<br>Sukhpal Singh (Registration No. - 801131024) |
| 3. Non-Conformance in software: - There is one small fault in calculating the distance between two towers and that is when we select the towers on map then tower images disappears from map screen. |
| 4. Non-Conformance in work products:-<br>i) Role name of different test engineers is not mentioned clearly. |

5. Grade of project:- 4.9/5.0
6. Remarks:-
i) The software system is in good working condition and satisfies quality requirements which are implicit with the domain of software.
ii) This software is confined to a single user (site engineer), so when to maintain dB of whole country with different states and cities, this software should be extended.
7. Auditors:-

| Name | Registration No. | Signature |
|---|---|---|
| VaibhavAggarwal | 801131028 | |
| Arvind Sharma | 801131007 | |

## IX. COMMUNICATION MANAGEMENT

Communications management is the methodicalarrangement, executing, observing, and reconsideration of all the networks of message within an organization, and among organizations; it also contains the organization and broadcasting of new communication instructionsassociated with an organization, network, or communications tools. Features of communications management comprise developing commercial communication policies, designing internal and external communications commands, and handling the movement of information, comprising online communication. New expertise forces persistentimprovement on the slice of communications administrators.

### A. Communication Management Plan

This is the manuscript that designates the communication anticipations, requirements, and tactics for the project. It stipulates what information will be transferred, when and how it will be conversed, and who will communicate it and to whom. It comprises: Communication desires of the project shareholders, Information to be transferred: content, format, and level of detail, who will transfer the information, who will accept it, and why, the person accountable for sanctioning the release of trustworthy information, Approaches of communication that will be used, such as e-mail, demonstrationand Communication restrictions. The communication management plan is described in Table X.

TABLE X
COMMUNICATION MANAGEMENT PLAN

| Name | Position | Internal/ External | Level of Interest | Level of Influence |
|---|---|---|---|---|
| Harinder Singh | Developer | Internal | High | Medium |
| Sukhpal Singh | Developer | Internal | High | Medium |
| Sukhwinder Singh | Telecom Company | External | Medium | High |
| Gurneet Singh | Site Engineer | External | Medium | High |
| Dr. Inderveer Chana | Project Manager | Internal | Medium | High |

### B. Stakeholder communication Analysis

Stakeholder communications investigation is available that may support the project management team to develop active communications. The stakeholder communication analysis is described in Table XI.

TABLE XI
STAKEHOLDER COMMUNICATION ANALYSIS

| Stakeholder | Document Name | Document Format | Contact Person |
|---|---|---|---|
| Customer Management | Project Charter | Hard Copy | Sukhwinder Singh |
| Customer Business and Technical Staff (Site Engineer) | SRS | Hard Copy | Gurneet Singh |
| Project Manager (Internal Management) | Weekly Status Report | Hard Copy | Dr. Inderveer Chana |
| Internal Business and technical staff (developer) | Weekly Status Report | e-mail | Harinder Singh & Sukhpal Singh |

## X. RISK MANAGEMENT

Risk management is the identification, valuation, and ranking of riskstrailed by synchronized and inexpensive application of resources to reduce, observer, and control the likelihood and influence of unsuccessfulactions or to maximize the apprehension of occasions.

### A. Risk Management Plan

Risk Management Plan is a manuscript that a project manager organizes to foreknow risks, guesstimateinfluences, and describecomebacks to problems. It also encompasses a risk assessment matrix. The risk management plan is described in Table XII.

TABLE XII
RISK MANAGEMENT PLAN

| Risk_Id | Risk Name | Category | Probability | Impact |
|---|---|---|---|---|
| R-101 | RAD Process Model proceed in uncontrolled manner | Project | 35% | 1 |
| R-102 | Schedule Estimate may less | Project | 40% | 2 |
| R-103 | Cost Estimate may be low | Project | 40% | 3 |
| R-104 | Team Members may lack in Skills | Human Resource | 30% | 2 |
| R-105 | Lack in Hardware | Resource | 45% | 4 |

| R-106 | Customer May feel uncomfortable with excessive interaction | Project | 35% | 2 |
|---|---|---|---|---|
| R-107 | Internet Connection | Technical | 50% | 1 |

## XI. CONCLUSION AND FUTURE WORK

The case teaching method has been used effectively in many professions (law, medicine, and business) to teach about how to solve problems and make decisions, while dealing with genuine circumstances, working with real-world restrictions, and engaging with both human and technical issues. Although the use of the case module in teaching software engineering has been restricted, the discipline is key candidate for such a procedure. The Case Study Project described in this paper has the objective of building a framework for using the case module for teaching software engineering. The goal of the OTPS software system is to provide a single comprehensive and complete example of the engineering of a software product. In addition, the project provides case modules (mini case studies), which can be used to teach various software engineering topics.

Future work will consist of testing of the current case study materials and ongoing development of the case study for subsequent phases and distribute the complete project in different software engineers. It is envisioned that the project will take about three years to complete.

## ACKNOWLEDGMENT

We would like to thank Dr. Inderveer Chana and Ms.Ashima Singh for helping and providing support in overcoming various technical obstacles in designing and implementation.We are also thankful to Dr.Maninder Singh, Head, Computer Science and Engineering Department, Thapar University, Patiala for his motivation, kind help and cooperation.

## REFERENCES

[1] Merriam, Sharan B. Qualitative Research and Case Study Applications in Education. Revised and Expanded from" Case Study Research in Education.".Jossey-Bass Publishers, 350 Sansome St, San Francisco, CA 94104, 1998.
[2] Merriam, Sharan B. Case Study Research in Education. A Qualitative Approach. Jossey-Bass Inc., Publishers, PO Box 44305, San Francisco, CA 94144-4305, 1988.
[3] Stake, Robert E. "The art of case study research." (1995).
[4] Eisenhardt, Kathleen M. "Building theories from case study research." Academy of management review (1989): 532-550.
[5] Gerring, John. Case study research: principles and practices. Cambridge: Cambridge University Press, 2007.
[6] Feagin, Joe R., Anthony M. Orum, and Gideon Sjoberg. Case for the case study. UNC Press Books, 1991.
[7] Boehm, Barry W. "Software engineering economics." Software Engineering, IEEE Transactions on 1 (1984): 4-21.
[8] Abran, Alain, J. Moore, P. Bourque, R. L. Dupuis, and L. Tripp. Guide to the Software Engineering Body of Knowledge–SWEBOK, trial version. IEEE-Computer Society Press, 2001.
[9] Pressman, Roger S., and Darrel Ince. Software engineering: a practitioner's approach. Vol. 5. New York: McGraw-hill, 1992.
[10] Davis, Alan M., Edward H. Bersoff, and Edward R. Comer. "A strategy for comparing alternative software development life cycle models." Software Engineering, IEEE Transactions on 14, no. 10 (1988): 1453-1461.
[11] Jacobson, Ivar, Grady Booch, and James Rumbaugh. "The Unified Software Development Process–The complete guide to the Unified Process from the original designers." Rational Software Corporation, US (1999).
[12] Nurmuliani, N., DidarZowghi, and S. Powell. "Analysis of requirements volatility during software development life cycle." In Software Engineering Conference, 2004. Proceedings. 2004 Australian, pp. 28-37. IEEE, 2004.
[13] Karolak, Dale Walter. Global software development: managing virtual teams and environments. IEEE Computer Society Press, 1999.
[14] Highsmith, Jim, and Alistair Cockburn. "Agile software development: The business of innovation." Computer 34, no. 9 (2001): 120-127.
[15] Hughes, Bob, and Mike Cotterell. Software project management. Tata McGraw-Hill Education, 2002.
[16] Henry, Joel. Software project management. Pearson, 2004.
[17] Atlee, Joanne M., T. C. Lethbridge, A. Sobel, J. B. Thompson, and R. J. LeBlanc Jr. "Software engineering 2004: ACM/IEEE-CS guidelines for undergraduate programs in software engineering." In Software Engineering, 2005. ICSE 2005. Proceedings. 27th International Conference on, pp. 623-624. IEEE, 2005.